\documentclass[conference]{IEEEtran}
\usepackage[numbers]{natbib}
\usepackage[T1]{fontenc}
\usepackage[utf8]{inputenc}
\usepackage{amsmath,amssymb,amsfonts}
\usepackage[inline]{enumitem}
\usepackage{algorithmic}
\usepackage{graphicx}
\usepackage{xcolor}

\begin{document}

\title{A Decade of Research for Image Compression In Multimedia Laboratory*\\
{\footnotesize \textsuperscript{*}Note: The authors are equally contributed in the current research}

}

\author{\IEEEauthorblockN{1\textsuperscript{st} Shahrokh Paravarzar}
\IEEEauthorblockA{\textit{dept. of Computing Science} \\
\textit{University of Alberta}\\
Edmonton, Alberta-Canada \\
Paravarz@Alberta.ca}
\and
\IEEEauthorblockN{2\textsuperscript{nd} Javaneh Alavi}
\IEEEauthorblockA{\textit{dept. of Computing Science} \\
\textit{University of Alberta}\\
Edmonton, Alberta-Canada \\
Javaneh@Alberta.ca}
\and
}
\maketitle

\begin{abstract}
With the advancement of technology, we have supercomputers with high processing power and affordable prices. In addition, using multimedia expanded all around the world. This caused a vast use of images and videos in different fields. As this kind of data consists of a large amount of information, there is a need to use compression methods to store, manage or transfer them better and faster. One effective technique, which was introduced is variable resolution. This technique stimulates human vision and divides regions in pictures into two different parts, including the area of interest that needs more detail and periphery parts with less detail. This results in better compression. The variable resolution was used for image, video, and 3D motion data compression. This paper investigates the mentioned technique and some other research in this regard.
\end{abstract}

\begin{IEEEkeywords}
image compression, variable resolution, human vision, videoconferencing
\end{IEEEkeywords}

\section{Introduction}
The utilization of the data compression techniques commenced from the first entries of the computers. Initial methods focused on text and program compression. However, nowadays by broad usage of still and moving images, there is a demand for applying special new compression techniques for this kind of data. Image compression methods can be beneficial in different fields such as the publishing industry, image, and movies archival systems, videoconferencing, and medical image storing. The aim of compression is to represent the same information with a smaller size. Two main methods for image compression are lossless and lossy techniques. In lossless methods, compression will not affect the data, which means the data will store in a different format, but pixel values will remain the same. In contrast, lossy techniques modify the data, reduce data size and at the same time produce a better compression ratio~\citet{basu1998enhancing}. There are various methods for solving different problems in the area of image compression. The aim of these techniques is to find more appropriate methods, which are faster, use less bandwidth, and have acceptable quality. One of the helpful approaches is to apply the variable resolution concept for image compression. In this approach, interesting regions of images (fovea) were represented with more detail while the other parts (periphery) were shown with less detail. There are some advantages to this method such as fast performance and Guaranteed compression ratio. Moreover, after using variable resolution (VR) compression the resulted images can be compressed more by applying other methods~\citet{basu1998enhancing}. The variable resolution techniques that align with other image and video compression processes are introduced in this paper.

\section{Review on Literature}
\citet{cheng2015perceptually} purposed the Linear Set Partitioning In Hierarchical Trees (LSPIHT) for motion capture data compression. The motion capture data compression is required to display animations in a real-time interactive transmission. The proposed method aims to use the inter and intera-channel for the high temporal and spatial correlated data to reduce the bandwidth used to animate these physical animations. Their contribution of research was aiming to reduce the bandwidth bu allocating the exact amount, make shorter the encoding and decoding process while preserving the scalibility. The purposed methodology preserve the efficient encoding for both short and long clips while it is providing a fast and accurate bandwidth allocation. The motion captured data are presented in hierarchical form while storing the joint rotation using the Euler rotation angle. The LSPIHT algorithm is similar to EZW that is applying to the multi-level wavelet decomposition. The bandwidth channel is defined by~\citet{cheng2015perceptually} as follows:
\begin{equation}
    B_{C} = \frac{F_{C}B_{T}}{\sum_{i=1}^{c} F_{i}}
\end{equation}
where $C$ is the channel, $B_{T}$ is the total bandwidth allocated to the current segment of a character motion and $C$ is the number of channels, $F_{C}$ is the importance score and $B_{C}$ is the allocated bandwidth. To obtain the semi-optimal solutions a series relative importance score of each channels used along with a heuristic methods. These factors are 1) body part score, 2) Position in the kinematic chain (depth factor) 3) Channel energy (variation). The character divide into six kinematic chains each with a score between 0 to 1 (called $S_k$) that $k$ is the kinematic chains.The position in the kinematic chain is can is presented on Eq.~\ref{refsh_1}:
\begin{equation}
    P(j_k) = \frac{h D(j_{k})+1}{h+1}
    \label{refsh_1}
\end{equation}
where $h$ is the maximum depth of the kinematic chain and $D(j_k)$ is the depth of the joint in the tree (each chain starts from depth 0). The last parameter is the channel variation which can be defined as the $i^{th}$ channel $(C_i)$ as
\begin{equation}
    E(C_i) = \frac{\sum_{f=1}^{W=1} (C_i(f+1))}{WD_{i}}
    \label{refsh_2}
\end{equation}
where $W$ is the number of frames (window size) in the current segment and $D_{i}$ is max of the $\sum_{f=1}^{W=1} (C_i(f+1))$ indicates the amount of variation (motion flow) in a channel. ~\citet{cheng2015perceptually} validated and compared their results with the Double Stimulus Continuous Quality Scale (DSCQS) method as dis- cussed in ITU-R recommendation BT.500 for salsa dancing, window washing and walking for a duration of 20 minutes for each cases using the indoor testing environment. T compared their results using Differential Mean Opinion Scores (DMOSs) vs. Compression Ratio (CR). Their purposed method were able to provide a for 60:1 with a DMOS of less that 0.2. The purposed compression method demonstrated that the compression time and decompression time per frame. The purposed approach achieve the low bandwidth and higher compression, fast compression and decompression and scalibility.\\
~\citet{kottayil2018adapting} studied the effect of the JPEG compression in the quality loss in 3D model textures. Their aim was to assess the perceptual quality during the transmission considering the bandwidth for 3D textural meshes. The textural meshes are representing in a uncompressed form. Their designed experience were consisted of number of references to preserve the self-consistency while setting the internal scale for distortion. Providing an inference would let the users to express their opinions while interacting with the subject as"Bad","Poor","Fair","Good" and "Excellent". Their study demonstrated that the average user score increased exponentially by increasing the perceived quality for each Q factor. They evaluate their model using the log function in form of $y = a /times log(b_cx)$ where $x$ is the $Q$ factor and $a$,$b$ and $c$ are the fitting parameters. The generic tex-mesh model the following formulation offered $t$ demonstrate the preserved quality for a geometry model level $g$ and the texture resolution $t$
\begin{equation}
\label{refsh3}
  Q(g,t) = \frac{1}{
  \frac{1}{m_(M-m)^{t}
   + (\frac{1}{m} - \frac{1}{m_(M-m)^{t}) (1-g)^{c} }}}  
\end{equation}
where $m$ and $M$ are the minimum and maximum bounds of quality, $g$ and $t$ are the graphical and texture components scaled into a $[0, 1]$ interval, and c is a constant. The authors claim that the Eq.~\ref{refsh3} can be reformulate by replacing the $t$ value as follows: 
\begin{equation}
    t^\prime = t \times (a^log(cx+b))\\
    a^= \frac{a}{5}
\end{equation}
~\citet{kottayil2018adapting} found that the user responses are related to the $Q$ factor and texture file size by Eqs.~\ref{refsh3} and ~\ref{refsh_2}.\\
\citet{sanchez2004prioritized} proposed the multiple region of interest encoding algorithm for JPEG2000 by prioritization method. In this approach the authors process the region of interest before the background and use the prioritization method to determine how much background should be included with the region of interest (ROI). The prioritization approach is done for each package using the distance from ROI by implementing to the foveated image. The packets will be gradually drop in a network congestion. the packets within the ROI recive the higher priority in contrast with their surrounding packets. A Gaussian distribution is used to prioritize the packages. The prioritization formula is as follows: 
\begin{equation}
\label{eqsh_6}
P_{m,l}(r) = \left\{\begin{matrix}
\frac{P_{ROI}}{L}.(L-l+1) & \mathrm{if packect} \in ROI\\ 
 \frac{P_{ROI}.2^(\frac{r}{R})^2}{L}& \mathrm{if packect} \notin ROI
\end{matrix}\right.
\end{equation}
where $P_{ROI}$ is the priority level assigned to the ROI, is the distance between the centers of the ROI and the region represented by packet,is the number of layers used to compress the image, and (the shape parameter) is defined as the radius at which the priority drops to one-half its maximum value. The approach is to prioritise the different layers. In comparison to the MAXSHIFT method the methodology introduced by~\citet{kottayil2018adapting} demonstrated better smoothly variate quality within an image.\\
~\citet{cheng2003parametric} presented the spatially variation sensing method to enhance the interactive and mesh transmission for online 3D applications. Some of the researchers implemented the mesh compression while merging the regions in a methods such as multi resolution LOD based methods, however in the current model the mesh and texture resolution vary continuously by transmission of the points of interest (foveae).~\citet{cheng2003parametric}express that the foveation is a reasonable method to transfer the data over a network with a proper quality while fast updating the network. In this method the vertices and textures near to the foveae will have a higher weight rather than the new fovea which are discarded and the new data will be transmitted if the user requires higher resolution from the mesh in a sub-sequence requests. The compression provide by the the server will be transmitted to the client side and then the client will reconstruct the image using the coarse model and the high-resolution data received. The foveation transform implemented using the CVR method considering the scaling factor ($s$) and distortion parameter ($\alpha$). These values will be define the quality of the surrounding area of the fovea.The presented method is fast and simple parametric method for concurrently mesh and texture compression.
~\citet{basu1996enhancing} presented the effect of fovea in stereo image compression algorithms using the spatial variation sensing method. The stereo images are taken from the two different perspectives. and the depth of each 3D points will be obtained from a relative displacement between its projections on the left and right hand side.~\citet{basu1996enhancing} took the advantage of visual human system and implement the stereo pair compression using the discete cosine transform (DCT) and VR transformation. The stereo pair of 320 by 240 pixel used for the experiment and the results evaluated for three case of uniform DCT, SVDCT and DCTSVS. The DCTSVS were the best method with respect to the processing time, however the DCT hardware become better and more common, so the SVDCT method can be used as a better choice.\\
~\citet{wiebe1997modelling} presented several modeling approaches which could represent the various visualisation system either human or animals. These modeling approaches would help the compression process to imitate the visual system behaviors based on foveated vision approaches. Single, multiple, moving and weighted fovea as well as visual streaks are presented by~\cite{wiebe1997modelling} demonstrated that the animal visual systems are the best system to perform complex manipulations and operations with the detailed optic data. Although the most compression methods does not considering the spatial relation of pixels in the image. The methods which simulate the biological system would perform better than the non-spatial methods.\\
~\citet{wiebe2001improving} assess the performance of the asynchronous transfer mode (ATM) using the variable resolution compression method. The asynchronous in ATM means ATM devices do not send and receive information at fixed speeds or using a timer, but instead negotiate transmission speeds based on hardware and information flow reliability.\\
The networks can be congested in presence of high traffic, the congested network could result in information lost gradually until total failure results in blindness, ATM networks allow data cell prioritization to control the loss. It means that in the presence of high traffic which cells can transfer and has got a high priority. This network has been simulated by the authors. The goal is the transmission of image video data to reduce the bandwidth required preserving reception quality. The parameters of compression should be set prior to sending the information with a specific bandwidth requirement, and any data reduction would could result in a significant dollar-cost saving depending on a compression ratio achieved.\\
ATM is based on fixed packet size of data or cells or fixed bit rate. Each cell is 53 bytes, 5 of each holds header information, ATM network requires that connection established prior to transmission, the user and the network will agree to the properties of connection, The QoS dictates the level of acceptable data loss, and the bandwidth limits bound to the data transmission rate


~\citet{basu1994videoconferencing} introduced different encoding techniques for images and explained one compression technique, which is variable resolution. This method was based on human vision. It was also mentioned in the paper that special part of pictures has more details (fovea) while the other parts which is periphery has less details. This effect can be achieved through using special cameras, which use log-polar coordinate or via a mathematical model. In this paper, a basic variable resolution model was proposed. However, a problem with this model is that the resulting images are not rectangular. Thus, the method was extended to solve this problem. There were two approaches for the extended model. First, “Modified Variable Resolution” maintains the isotropic properties, but it is complicated. Second, “Cartesian Variable Resolution” reduces isotropic accuracy, but it is less complex. Moreover, moving fovea and multiple foveae were described. When Compressing several images in a continuous sequence, the position of the fovea will change. This is called moving fovea and can be useful for videophone applications. In cases that we have more than one point of interest in an image, it is called multiple foveae. Multiple fovea relationships were defined as “cooperative” and “competitive”. At last, based on the variable resolution method, a videophone prototype was introduced.

\begin{figure}[ht]
    \centering
    \includegraphics[width=0.45\textwidth]{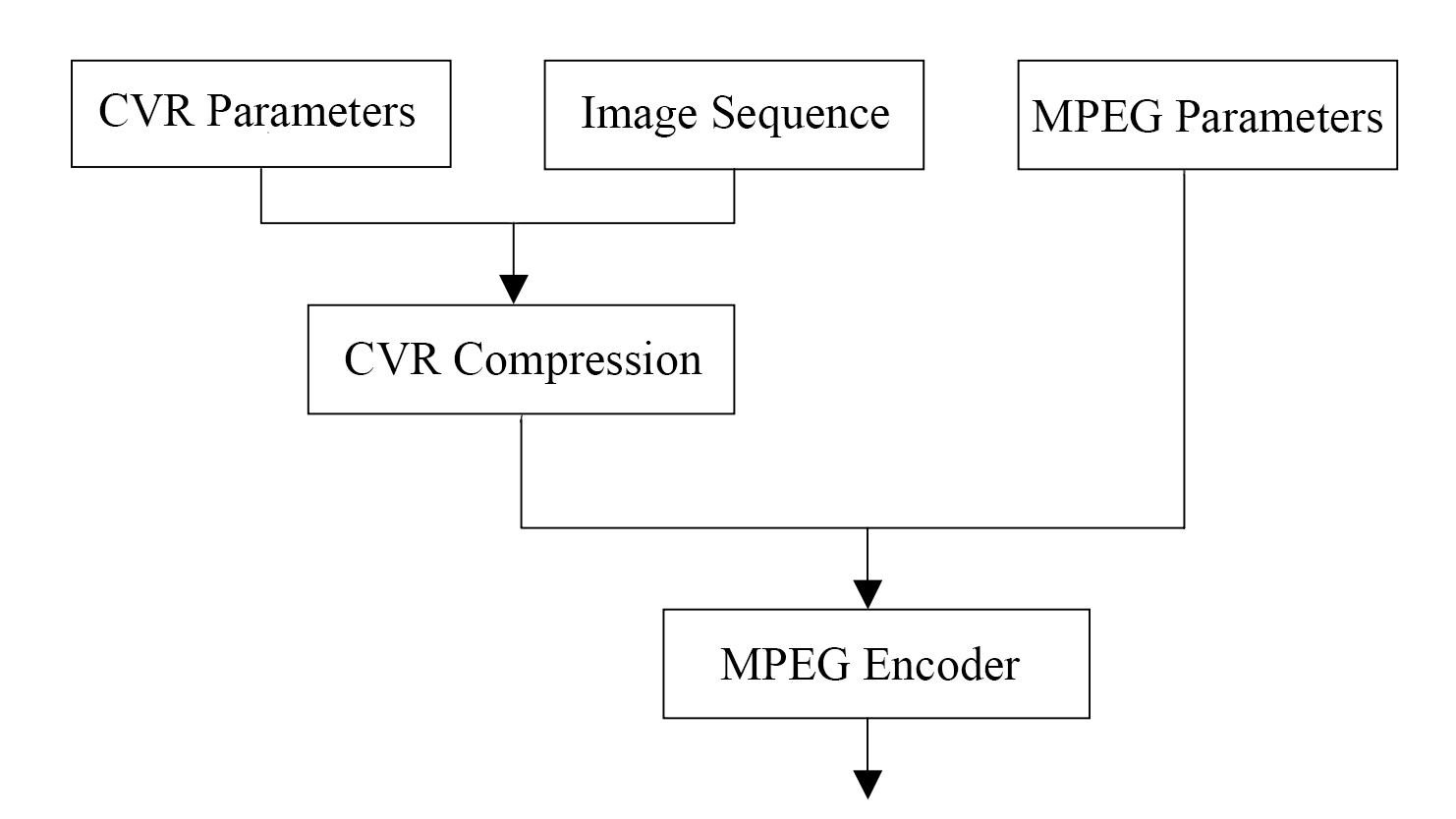}
    \caption{CVR/MPEG video compression~\cite{cheng2003qos}}
    \label{fig:1}
\end{figure}

\begin{figure}[ht]
    \centering
    \includegraphics[width=0.45\textwidth]{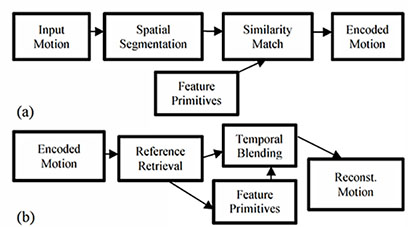}
    \caption{Processing pipeline of the proposed method: (a) 
encoding and (b) reconstruction.~\cite{firouzmanesh2013efficient}}
    \label{fig:2}
\end{figure}

~\citet{basu1993variable} recited that Images consisted of a large amount of information and are widely used in digital media. Therefore, in order to use digital images a kind of compression is required. This article suggested a variable resolution (VR) technique in this regard. This method focused on the interesting part of the images, which is called fovea. In this part, more detail is presented, and less detail will be shown in the surrounding part (periphery). This method advantageous are being fast, guarantees compression ratio and more compression can be created by using other methods. After introducing and using this technique, a comparison was made with different methods. Results showed that the original method can be enhanced with some more computation. Moreover, this method is used to establish a teleconferencing system. Fast compression requirement, no necessity for high quality, and having typically one important part (head) in a scene as fovea location, are some reasons for using VR compression.  Outcomes demonstrated the usefulness of the VR method for compression on machines with limited speed as well as allowed video conferencing with little hardware cost.\\
~\citet{cheng2003qos} expanded their past works on foveation. The advantages of this technique on MPEG compression over limited bandwidth such as the internet was investigated. In this approach, interesting regions in a video was transmitted with high resolution, while the other parts were transferred at low resolution. Continuously, spatially varying resolution was used. An advantage of this approach was that there will not be sudden resolution change at boundaries between the region of interest and the peripheral region. This paper suggested a combination of CVR compression technique and MPEG. The image was compressed by MPEG algorithm after the input image processed through CVR compression. For decompression first, MPEG decoding was used followed by inverse CVR (Fig~\ref{fig:1}). By using scaling and distortion parameters, MPEG videos were prepared by different compression ratios on the server. Then, the proper video was delivered depending on existing network resources and user requirements. By statistical approach, network bandwidth was optimally observed. This method achieved Quality-of-Service delivery on the application layer without changing the network architecture. The results demonstrated the effectiveness of this method compared with traditional multiresolution MPEG.
As the use of the internet and multimedia is becoming more popular and interactive the need to transfer high-quality 3D objects is becoming more important. Traditional ways to generate 3D images like colored polygons and ray tracing are not realistic or demand a large amount of data loading. Therefore,~\citet{basu2002foveated} extended the use of varying sensing for 3D visualization. This method, which was previously defined for image compression, was used to improve interactive 3D visualization over limited bandwidth such as the internet. In this work an algorithm was implemented in JAVA3D. A foveating JPEG texture was combined with a level of detail representation in JAVA3D. Although 3D applications are becoming more important, the limited bandwidth is an issue. By using foveated 3D visualization, the region of interest has more details, and it will adapt to changing bandwidth. In this mechanism, different qualities of files were stored on the server and will be used depending on the quality that the user needs as well as existing bandwidth. First, the application will run on client-side with low resolution. Afterward, the user selects an area of interest and the fovea information is transmitted. The server generates and compresses the texture image based on the fovea and look-up table. The result is sent to the client and the client reconstructs the 3D object. In this way, decreasing the demand for data retrieval.In this paper was also mentioned that a combination of foveation-DCT method can be useful for online 3D applications. In 3D scenes near parts are more important and it is reasonable to use foveated technique. The other advantage of this method is that multiple foveae selections can be gathered to decrease delivery time. At last, a user interface regarding this method was developed.\\
In another paper by~\citet{firouzmanesh2011perceptually} a lossy compression technique and algorithm was proposed for motion data by considering human perception. This method would not reduce the quality of animation a lot via considering some important factors. This method is faster than similar methods and suitable for situations, where processing power and memory capacity is limited. This paper reviews some motion data compression techniques and study results on human perception of animation and introduced a new method that aimed to reduce processing resources for encoding and decoding as well as integrate human perceptual factors of animation. The idea is to choose different coefficients for different channels of data considering the importance of the channel on observed quality of motion. Global optimization method is not used. Length of the bone connected to the joint and variation in rotation were two important factors that were taken into consideration. This method is fast compared to other similar methods. At least a 25:1 compression ratio was achieved without significantly influencing the results This is useful for a mobile environment.\\
~\citet{firouzmanesh2013perceptually} mentioned the transmission problem of real-time motion data in an online environment and proposed a lossy compression method that was efficient and considered human perception. The method used incremental encoding and a database of motion primitives for each key point. This database makes the applying of repetitive motions easier. Databases have been used in previous works. However, there was a need to take a new approach to apply that for real-time dynamic motions. Some motions, which may be unpredictable and having a complete predefined database, are not achievable. Therefore, compression was divided into offline and online parts to reduce the online processing time. For the offline phase, a database was made based on training motion data and saved at the rendering site. In the online part, they compressed every segment of the motion. Then, referenced every segment to its closest primitive in the data set, if the difference between segment and closest primitive was under a certain threshold. Otherwise, incremental encoding was applied to transmit the compressed difference. The method was implemented, and one important feature was controlling the trade-off between quality and bandwidth by one parameter. Results show that this method achieved a higher compression rate with acceptable quality compared with other methods in the state-of-the-art. The processing time was reduced, and real-time performance was also archived.\\
~\citet{firouzmanesh2013efficient} combined spatial segmentation and temporal blending on rhythmic motion data and explored its capability to improve the compression ratio and quality of render. Main primitive movies that were needed for compression stored in a database. They were generated via segmenting the sample motions. Thereafter, it was compressed by encoding it thorough changing each segment with the best matching primitive in the database. After that, by blending the primitives, the decoding was done (Fig~\ref{fig:2}). Results showed the usefulness of this method mainly for long sequences. It decreased visual artifacts and produced suitable quality for output animation.
~\citet{yee2017medical} expressed that For medical images compression is needed for better storing, managing, and transferring. In this paper, it is mentioned that some parts of images may be more important than other parts. Therefore, in compression techniques, important parts of images should be reconstructed after decompression with more details. In this regard, this study used Better Portable Graphics (BPG) based on High-Efficiency Video Coding (HEVC) for compressing medical images. This method consists of the following parts. First, segmenting a medical image into two regions: region of interest and non-region of interest. After that, applying lossless BPG to the region of interest and lossy BPG for the rest part. Eventually, the two parts were combined to construct the complete compressed image. Evaluations related to this technique showed 10-25\% improvement in compression rate compared to older techniques.
\section{conclusion}
In this paper different image compression techniques and their application in variety areas were presented including image or video compression, 3D visualization, motion capture data, and medical image compression. Among these techniques, variable resolution was one of the beneficial approaches for image compression, where speed and hardware resources are limited. This method was especially useful for teleconferencing and videoconferencing. In addition, based on this idea, a videophone prototype was implemented. The use of this method was also extended for 3D visualization. Overall, some image compression techniques, which were summarized in this paper, can inspire more investigation on this interesting and demanding field of study.
\section*{Acknowledgment}
We gratefully appreciate the assistant of Dr. Basu  for providing guidance and helpful advice.

\bibliography{Bibliography.bib}
\bibliographystyle{IEEEtranN}

\end{document}